\documentclass[traditabstract]{aa}
\usepackage{epsfig}    
\usepackage{lscape}
\usepackage{natbib}
\bibpunct{(}{)}{;}{a}{}{,}    %% natbib format like A&A and ApJ
\usepackage{graphics}
\usepackage{graphicx,graphics}
\usepackage{color} 
\usepackage{times}
\usepackage{amsmath}
\usepackage{amssymb}
\begin{document}

\title{Spiral galaxies with unusual azimuthal asymmetry in oxygen abundance} 
      
\author{
         L.~S.~Pilyugin\inst{\ref{ITPA},\ref{MAO}}            \and 
         G.~Tautvai\v{s}ien\.{e}\inst{\ref{ITPA}}     
}
\institute{Institute of Theoretical Physics and Astronomy, Vilnius University, Sauletekio av. 3, 10257, Vilnius, Lithuania \label{ITPA} 
\and
  Main Astronomical Observatory, National Academy of Sciences of Ukraine, 27 Akademika Zabolotnoho St, 03680, Kiev, Ukraine \label{MAO}
  }

\abstract 
{
  We considered five galaxies in the Mapping Nearby Galaxies at the Apache Point Observatory (MaNGA) survey that show distinct azimuthal asymmetry in the abundance, 
  in the sense that in the inner part (more than half of the optical radius, $R_{25}$) of each galaxy there is a sector-like (up to semi-circle) region where the oxygen abundances 
  (O/H)$_{\rm h}$ are higher than the abundances (O/H)$_{\rm l}$ in other sectors. M-11761-12705 is a massive galaxy with a stellar mass of log($M_{\star}/M_{\sun}$) = 11.6; 
  the masses of four other galaxies (M-8450-12701, M-8546-12704, M-8561-12701, M-9500-09102) are moderate: 10.1 $\protect\la$ log($M_{\star}/M_{\sun}$) $\protect\la$ 10.4.
  Abundances within both high- and low-metallicity regions show flat radial gradients (the abundances are at nearly constant levels). The histogram for the spaxel abundances
  demonstrates two distinct peaks, the difference  between the (O/H)$_{\rm h}$ and the (O/H)$_{\rm l}$ abundances are of 0.06 -- 0.08~dex. The high-metallicity regions are located
  in the O/H -- N/O diagram closer to the lower envelope of the band than the low-metallicity regions. The abundance properties in the massive galaxy M-11761-12705 can be
  explained by the low-metallicity gas infall onto the galaxy and subsequent episode of high star formation rate (starburst) in the diluted interstellar medium occurring between
  20~Myr $\protect\la$ $t$ $\protect\la$ 50~Myr ago. For moderate-mass galaxies, the higher oxygen abundance in the high-metallicity region and its shift towards the lower envelope
  in the N/O -- O/H diagram compared to the low-metallicity  region can be explained in one of two ways: either the starburst in the high-metallicity region occurred several
  dozens of Myr ago, or the star formation in the galaxy is accompanied  by galactic winds, and the region evolved with the lower efficiency of the enriched galactic winds shows
  higher metallicity.  Two galaxies of our sample (M-8546-12704 and M-11761-12705) are members of galaxy pairs. However, the asymmetry parameter, $A$, quantifying the asymmetry of
  a light distribution across the galaxy, is above the canonical threshold ($A$ = 0.35) between non-interacting and interacting galaxies in the massive galaxy M-11761-12705 only.
  The values of the $A$ parameter in four moderate-mass galaxies (including a member of the galaxy pair) are below the threshold value. 
}


\titlerunning{MaNGA galaxies with unusual azimuthal asymmetry in oxygen abundance}
\authorrunning{Pilyugin and Tautvai{\v s}ien\.{e}}

\maketitle

\section{Introduction}
%=====================

The discs of spiral galaxies have long been known to show negative radial abundance gradients in the sense that the abundance is higher at the centre and decreases with the galactocentric
distance \citep{Searle1971,Smith1975}. The distribution of nebular abundances across the disc of a galaxy is specified by a relation between the oxygen abundance, O/H, and the galactocentric
distance, $R$. Relations of this type were determined for many galaxies by different authors \citep[][among many others]{VilaCostas1992,Zaritsky1994,Pilyugin2004,Pilyugin2007,Pilyugin2014,
Pilyugin2019,Gusev2012,Sanchez2014,Ho2015,SanchezMenguiano2016,Belfiore2017,SanchezMenguiano2018,Kreckel2019,Berg2020,Pilyugin2024}.

It is a common practice to derive the relation between the oxygen abundance, O/H, and the galactocentric distance, $R,$  assuming that the abundance within the disc is symmetric.
Numerical simulations predict azimuthal variations in abundances \citep[e.g.][]{Grand2016,Bellardini2021,Bellardini2022,Orr2023}. 
Integral field unit (IFU) spectroscopy of galaxies provides the possibility of constructing the abundance map and  investigating the azimuthal variations in metallicity within the disc.
The uncertainties in the abundance determinations (e.g. due to uncertainties in the emission lines measurements,  uncertainties in the calibrations used in the abundance determinations,
and contamination of the  H\,{\sc ii} region spectra by the non-stellar (shochs) ionising radiation) can make a contribution to the scatter in the measured oxygen abundances at a given radius. 
\citet{Sanchez2015} analysed the abundance distribution across the galaxy NGC~6754 using the IFU spectroscopy obtained with the Multi Unit Spectroscopic Explorer (MUSE)   
and found evidence of an azimuthal variation in the oxygen abundance with an amplitude of $\sim$0.05~dex. \citet{Zinchenko2016} constructed maps of the oxygen abundance across
the discs of 88 galaxies using observations obtained by the CALIFA survey (Calar Alto Legacy Integral Field Area survey; \citealt{Sanchez2012}). They found that the scatter in oxygen
abundances around the abundance gradient is within 0.05~dex. \citet{Ho2018} found 0.06 dex azimuthal variations in the oxygen abundance  in NGC~2997. At a given radial distance, the
oxygen abundances are higher in the spiral arms and lower in the inter-arm regions. Mapping both the radial and azimuthal variations in metallicity across a sample of galaxies is a
key science goal of the PHANGS (Physics at High Angular Resolution in Nearby GalaxieS) project. \citet{Kreckel2019} mapped the two-dimensional distribution of the gas-phase oxygen abundances
for 7138  H\,{\sc ii} regions across the discs of eight nearby galaxies using Very Large Telescope/Multi Unit Spectroscopic Explorer (MUSE) optical integral field spectroscopy obtained
within the PHANGS project. They found a very small root-mean-square scatter in residual metallicity (0.03--0.05~dex) after subtracting the radial gradients.   
\citet{Grasha2022} determined oxygen abundances in a spatially resolved  H\,{\sc ii} regions in six local star-forming and face-on spiral galaxies. They found a negative relation between
the gas-phase oxygen abundance and galactocentric distance. All metallicity radial profiles were well approximated with a simple linear metallicity gradient with a small (mean 0.03 dex) scatter.
NGC 1566 was an exception, with a significant scatter and a hint of flattening of the radial metallicity profile beyond 15 kpc from the galactic centre.
\citet{Bellardini2021}  examined variations of gas-phase  oxygen abundances using simulations for Milky Way and M31-mass galaxies across their formation histories at $z \lid$ 1.5 (look-back time
$\lid$ 9.4~Gyr). They found that over time the azimuthal variations become weaker; azimuthal variations are typically 0.14 dex at $z$ = 1 and become as small as 0.05 dex at $z$ = 0.

In some cases, the azimuthal variations in the oxygen abundance can be attributed to the spiral arms \citep{SanchezMenguiano2017,Sakhibov2018,Ho2018,Kreckel2019,SanchezMenguiano2020}.
However, \citet{Kreckel2019} found, for eight PHANGS galaxies, that azimuthal variations are not always obviously associated with the spiral pattern. In all cases, the azimuthal trends are more
pronounced on only one spiral arm.  \citet{Williams2022} examined the 2D variations in metals across the discs of 19 PHANGS galaxies and found no evidence that spiral arms are enriched
compared to the disc. \citet{SanchezMenguiano2020} found the presence of more metal-rich  H\,{\sc ii} regions in the spiral arms with respect to the corresponding interarm regions for
45--65 \%\ of galaxies while 5--20 \%\ of galaxies display the opposite trend; i.e. more metal-poor H\,{\sc ii} regions are in the spiral arms compared to the inter-arm regions.

\citet{Kreckel2019} noted that  H\,{\sc ii} regions with enhanced or reduced metallicity are located across the full disc. \citet{Grasha2022} also found that  H\,{\sc ii} regions with enhanced
and reduced abundances are distributed throughout the whole disc. Thus,  spiral galaxies reveal an azimuthal variation in the oxygen abundance with an amplitude within $\sim$0.05~dex, and
regions with enhanced and reduced abundances are distributed throughout the whole disc.

We found five galaxies   in the Mapping Nearby Galaxies at Apache Point Observatory survey (MaNGA, \citet{Bundy2015}) with unusual abundance distributions. In the inner (more than half
of the optical radius) part of each galaxy, there is a large, sector-like (up to a semi-circle) region where the oxygen abundances are appreciably higher ($\ga$ 0.06~dex) than in another
sector, that is, the difference between the oxygen abundances in those sectors exceed the typical value of the scatter around the O/H -- $R$ relations and an amplitude of the reported
azimuthal variations in the oxygen abundance \citep[e.g.][]{Sanchez2012,Zinchenko2016,Kreckel2019,Grasha2022}.
An azimuthal asymmetry of this kind cannot be attributed to the spiral arms. The study of such galaxies can shed light on some aspects of the (chemical) evolution of galaxies. 
This paper is organised as follows: the oxygen abundance and other characteristics of selected galaxies are described in Sect.~2; the analysis of the parameters of these galaxies and
the discussion of possible way(s) of the formation of the abundance distributions are given is in Sect.~3; and Sect.~4 contains a brief conclusion.

\section{Data and galaxy sample}
%=====================

\subsection{Data}
%=====================

Our investigation is based on galaxies from the MaNGA survey (\citealt{Bundy2015}). The completed observations of MaNGA galaxies were included in Data Release 17 \citep{Abdurrouf2022}.
The MaNGA data products were also revised for all the observations previously released in DR15 and prior releases (e.g. the flux calibration was updated). The emission-line parameters of
the spaxel spectra for our sample of galaxies are available from the MaNGA Data Analysis Pipeline (DAP) measurements. We derived the characteristics for a sample of 430 galaxies
using the latest version of the DAP measurements (manga-n-n-MAPS-SPX-MILESHC-MASTARSSP.fits.gz\footnote{https://data.sdss.org/sas/dr17/manga/spectro/analysis/v3\_1\_1/3.1.0/SPX-MILESHC-MASTARSSP/}
for the spectral data and the Data Reduction Pipeline (DRP) measurements manga-n-n-LOGCUBE.fits.gz\footnote{https://dr17.sdss.org/sas/dr17/manga/spectro/redux/v3\_1\_1/} for the photometric data). 

We selected a sample of galaxies using the following criteria. First, we chose the discy galaxies for which the curves of iso-velocities in the measured line-of-sight velocity fields
resemble a set of parabola-like curves (hourglass appearance of the rotation disc). This condition provides a possibility to determine the geometric parameters of a galaxy,
which are necessary for determining the galactocentric distances of individual spaxels and constructing radial distributions of the characteristics across the galaxy. Using this criterion,
we also rejected strongly interacting and merging galaxies in which the line-of-sight velocity field is distorted to such an extent that the determination of the geometrical angles and
rotation curve is impossible. Second, galaxies with an inclination angle larger than $\sim$70$\degr$ were rejected because the fit of the H$\alpha$ velocity field in such galaxies can
produce unrealistic values of the inclination angle \citep{Epinat2008}, and, consequently, the estimated galactocentric distances of the spaxels can involve large uncertainties; moreover,
the interpretation of the abundance in individual spaxel is not beyond the question since the spaxel spectra involves radiations that originated in volumes at different galactocentric
distances along the line of sight. Third, we considered MaNGA galaxies mapped with 91 and 127 fibre IFUs, covering 27$\farcs$5 and 32$\farcs$5 on the sky (with a large number of spaxels
over the galaxy image). We only chose galaxies for which the spaxels with measured emission lines are well distributed across galactic discs and cover more than $\sim 0.8~R_{25}$.

The determination of the geometrical galaxy parameters (the coordinates of the rotation centre, the position angle of the major kinematic axis, and the inclination angle), rotation curve,
surface-brightness profile, optical radius of the galaxy, radial distributions of the oxygen abundance, and other characteristics are described in detail in our previous papers
\citep{Pilyugin2018,Pilyugin2019,Pilyugin2020,Pilyugin2021}. In brief, the geometrical parameters of the galaxy and the rotation curve are derived by best fitting the observed line-of-sight
velocity field using an iterative procedure \citep{Pilyugin2019}. The galactocentric distances of the spaxels determined with the obtained coordinates of the rotation centre of the galaxy
and kinematic angles were used to construct the radial distributions of the oxygen and nitrogen abundances and other characteristics across the disc of the galaxy. The measurements in
the SDSS filters $g$ and $r$ for each spaxel were converted into $B$-band magnitudes following \citet{Pilyugin2018}.  The radial surface-brightness distribution was fitted within the
optical radius by a broken exponential profile for the disc and by a general S\'{e}rsic profile for the bulge:
\begin{eqnarray}
       \begin{array}{lll}
I(r) & = & I_{\rm e}\exp \{-b_{\rm n}[(r/r_{\rm e})^{\rm 1/n} - 1]\} \\
     & + & I_{\rm 0,inner}\exp(-r/h_{\rm inner}) \;\;\;\; if \;\;\; r < R^{*} ,                        \\
     & = & I_{\rm e}\exp \{-b_{\rm n}[(r/r_{\rm e})^{\rm 1/n} - 1]\} \\
     & + & I_{\rm 0,outer}\exp(-r/h_{\rm outer}) \;\;\;\; if \;\;\; r > R^{*}  .                        \\
     \end{array}
\label{equation:decomp}
\end{eqnarray}
Here, $R^{*}$ is the break radius, i.e. the radius at which the exponent changes.  The eight parameters, $I_{\rm e}$, $r_{\rm e}$, $n$, $(\Sigma_{\rm L})_{\rm 0,inner}$, $h_{\rm inner}$,
$(\Sigma_{\rm L})_{\rm 0,outer}$, $h_{\rm outer}$, and $R^{*}$, in the broken exponential disc were determined through the best fit to the observed surface-brightness profile. The value of
the isophotal radius, $R_{25,}$ was estimated using the fit corrected for the galaxy inclination.

The distances to the galaxies were taken from the NASA/IPAC Extragalactic Database ({\sc ned}).\footnote{{\sc ned} is operated by the Jet Populsion Laboratory at the California Institute 
  of Technology, under contract with NASA. {\tt http://ned.ipac.caltech.edu/}}  The {\sc ned} distances use flow corrections for Virgo, the Great Attractor, and Shapley Supercluster infall
(adopting a cosmological model with $H_{0} = 73$ km/s/Mpc, $\Omega_{\rm m} = 0.27$, and $\Omega_{\Lambda} = 0.73$). We adopted the spectroscopic stellar masses of the Sloan Digital Sky Survey
(SDSS) and BOSS \citep[BOSS stands for the Baryon Oscillation Spectroscopic Survey in SDSS-III, see][]{Dawson2013}. The spectroscopic masses were taken from the table
{\sc stellarMassPCAWiscBC03} and were determined using the Wisconsin method \citep{Chen2012} with the stellar population synthesis models from \citet{Bruzual2003}. The reported errors in
the values of the stellar mass are usually within 0.15 -- 0.2~dex. 

The measured line fluxes were corrected for the interstellar reddening using the reddening law of \citet{Cardelli1989} with $R_{\rm V}$ = 3.1. We classified the excitation of the spaxel
spectrum using its position on the  standard diagnostic Baldwin--Phillips--Terlevich (BPT) diagram:  [N\,{\sc ii}]$\lambda$6584/H$\alpha$ versus the [O\,{\sc iii}]$\lambda$5007/H$\beta$,
as suggested by \citet{Baldwin1981}. As in our previous studies \citep{Pilyugin2020,Pilyugin2021,Pilyugin2024}, the spectra located to the left (below) the demarcation line of
\citet{Kauffmann2003} are referred to as the SF-like or H\,{\sc ii} -region-like spectra; those located to the right of (above) the demarcation line of \citet{Kewley2001} are referred to
as the AGN-like spectra; and the spectra located between both demarcation lines are classified as intermediate (INT) spectra.

The oxygen abundances in spaxels with H\,{\sc ii} -region-like spectra were determined through the $R$ calibration from \citet{Pilyugin2016}. The calibration relation for H\,{\sc ii} regions
with log$N_{2} \ge -0.6$  (the upper branch) is
\begin{eqnarray}
       \begin{array}{lll}
     {\rm (O/H)}^{*}_{\rm R}  & = &   8.589 + 0.022 \, \log (R_{3}/R_{2}) + 0.399 \, \log N_{2}   \\  
                          & + &  (-0.137 + 0.164 \, \log (R_{3}/R_{2}) + 0.589 \log N_{2})   \\ 
                          & \times &  \log R_{2}   \\ 
     \end{array}
\label{equation:ohru}
,\end{eqnarray}
and the relation for H\,{\sc ii} regions with log$N_{2} < -0.6$ (the lower branch) is
\begin{eqnarray}
       \begin{array}{lll}
     {\rm (O/H)}^{*}_{\rm R}  & = &   7.932 + 0.944 \, \log (R_{3}/R_{2}) + 0.695 \, \log N_{2}   \\  
                          & + &  (0.970 - 0.291 \, \log (R_{3}/R_{2}) - 0.019 \log N_{2})   \\ 
                          & \times & \log R_{2}   \\ 
     \end{array}
\label{equation:ohrl}
,\end{eqnarray}
where (O/H)$^{*}_{\rm R}$ = 12 +log(O/H)$_{\rm R}$, and standard notations for the fluxes $R_2$  = $I_{\rm [O\,II] \lambda 3727+ \lambda 3729} /I_{{\rm H}\beta }$,
$N_2$  = $I_{\rm [N\,II] \lambda 6548+ \lambda 6584} /I_{{\rm H}\beta }$, and $R_3$  = $I_{{\rm [O\,III]} \lambda 4959+ \lambda 5007} /I_{{\rm H}\beta }$ are used. The nitrogen-to-oxygen ratios were estimated using
the corresponding calibration relation from  \citet{Pilyugin2016}:
\begin{eqnarray}
       \begin{array}{lll}
\log {\rm (N/O)}  & = &  -0.657 - 0.201 \, \log N_{2}  \\  
                  & + & (0.742 -0.075 \, \log N_{2}) \times \log(N_{2}/R_{2}) . \\ 
     \end{array}
\label{equation:nolin}
\end{eqnarray}

\subsection{Galaxy sample}

\begin{figure*}
\resizebox{1.00\hsize}{!}{\includegraphics[angle=000]{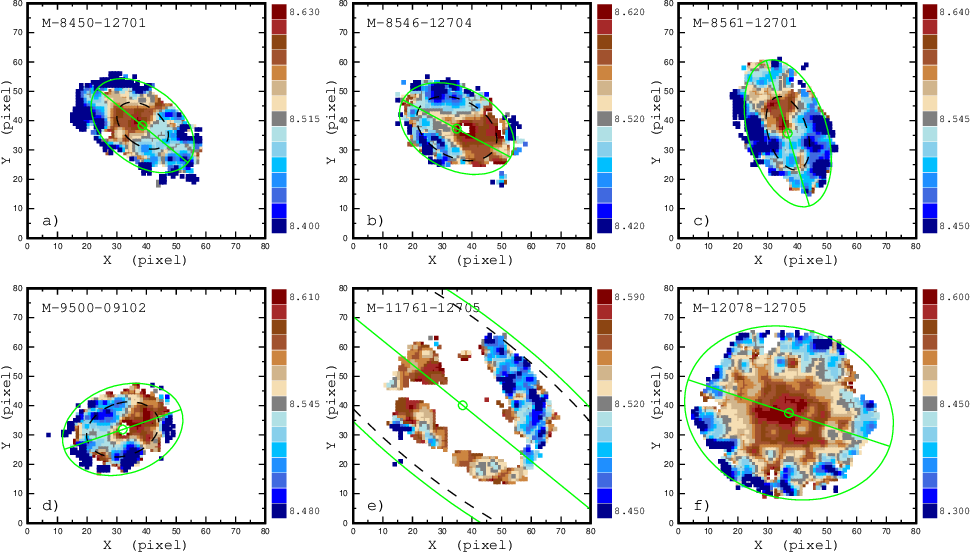}}
\caption{
  Oxygen abundance maps for our sample of selected target MaNGA galaxies with azimuthal asymmetry in the abundance distribution across the galaxy and control galaxy.
Each panel shows the distribution of the oxygen abundance across the image of the galaxy in sky coordinates (pixels). The value of the oxygen abundance in the spaxel is colour-coded.
The circle shows the kinematic centre of the galaxy, the line indicates the position of the major kinematic axis of the galaxy, and the solid ellipse is the 
optical radius. The dashed ellipse shows the area considered in the current study.
 {\em Panels} {\bf a, b, c, d, and e} show the target galaxies;  {\em panel} {\bf f} displays the control galaxy. 
}
\label{figure:oh-map}
\end{figure*}

By visual inspection of the oxygen abundance maps for 430 galaxies,  we find five galaxies that show distinct azimuthal asymmetry in the abundance in the sense that in the inner part
(more than a half of the optical radius $R_{25}$ of the galaxy) of each galaxy there is a sector-like (semi-circle in one case) region where the oxygen abundance (O/H)$_{\rm h}$ is higher
than the abundance (O/H)$_{\rm l}$ in other sector: M-8450-12701 (with a region of azimuthal O/H asymmetry within 0.5~$R_{25}$), M-8546-12704 (with a region of azimuthal O/H asymmetry within
0.7~$R_{25}$), M-8561-12701 (with a region of azimuthal O/H asymmetry within 0.5~$R_{25}$), M-9500-09102 (with a region of azimuthal O/H asymmetry within 0.6~$R_{25}$),   and
M-11761-12705 (with a region of azimuthal O/H asymmetry within 0.9~$R_{25}$). Fig.~\ref{figure:oh-map} shows oxygen abundance maps for the selected target galaxies with azimuthal O/H asymmetry.
The galaxy M-12078-12705 with the more or less symmetrical oxygen abundance (control galaxy) is shown in panel (f) of Fig.~\ref{figure:oh-map} for comparison. The characteristics of the
target galaxies are reported in  Table~\ref{table:general}. These galaxies were investigated in the current study.

\begin{table*}
\caption[]{\label{table:general}
  Characteristics\tablefootmark{a} of the MaNGA galaxies of our sample. 
}
\begin{center}
\begin{tabular}{ccccccccccccll} \hline \hline
Galaxy                  &
$d$                     &
log\,$M_{\star}$          &
$R_{25}$                 &
(O/H)$_{\rm h}$           &
(O/H)$_{\rm l}$           &
(O/H)$_{0}$              &
grad                    &
d(O/H)$_{\rm asym}$       &
SFR                     &
environ-               \\
ID                     &
Mpc                    &
$M_{\sun}$              &
Kpc                    &
12+log(O/H)            &
12+log(O/H)            &
12+log(O/H)            &
dex/R$_{25}$            &
dex                    &
$M_{\sun}$/year         &
ment                   \\
\hline
(1)                    &
(2)                    &
(3)                    &
(4)                    &
(5)                    &
(6)                    &
(7)                    &
(8)                    &
(9)                    &
(10)                   &
(11)                   \\
\hline
8450 12701  &  197.4  &  10.1  &   9.47  & 8.586 &  8.501 & 8.586\tablefootmark{b}
                                                                        & -0.098 &  0.041       & -0.304  & isolated   \\ 
  8546 12704  &  210.5  &  10.4  &  10.51  & 8.592 &  8.519 & 8.588     & -0.089 &  0.069       &  0.275  & multu      \\ 
  8561 12701  &  210.2  &  10.4  &  13.15  & 8.597 &  8.522 & 8.581     & -0.090 &  0.053       &  0.038  & isolated   \\ 
  9500 09102  &  193.7  &  10.2  &   9.77  & 8.597 &  8.534 & 8.580     & -0.062 &  0.042       &  0.083  & isolated   \\ 
 11761 12705  &  128.6  &  11.6  &  27.56  & 8.555 &  8.491 & 8.586     & -0.128 &  0.030       &  0.362  & multu      \\ 
                    \hline
\end{tabular}\\
\end{center}
\tablefoottext{a}{Distance to the galaxy $d$, stellar mass $M_{\star}$,  optical (or isophotal) radius $R_{25}$, oxygen abundances of the high (O/H)$_{\rm h}$ and the low (O/H)$_{\rm l}$ metallicity
  regions,  central oxygen abundance in the galaxy (O/H)$_{0}$,  slope of the metallicity gradient,  value of the global azimuthal asymmetry for the whole galaxy d(O/H)$_{\rm asym}$,
  current star formation rate (SFR), and  environment. \\
} 
\tablefoottext{b}{ The radial abundance distribution in M-8450-12701 is described by broken relation with the central abundance of (O/H)$_{0}$ = 8.586 and gradient of -0.098 dex~$R_{25}^{-1}$
  at $R/R_{25} < 0.77$ and (O/H)$_{0}$ = 8.829 and gradient of -0.415 dex~$R_{25}^{-1}$ at $R/R_{25} > 0.77$.
}  
\end{table*}

Galaxies M-8450-12701, M-8561-12701, and M-9500-09102 are isolated galaxies, the galaxy M-11761-12705 is a member of a pair of galaxies \citep{Tempel2018}. The M-8546-12704 is an isolated
galaxy according to \citet{Tempel2018}. However, the Sc galaxy (PGC 2412032, SBS 1552+524B, where SBS is the Second Byurakan Survey) is located at the distance of $\sim$2$\farcm$7 in the
sky plane corresponding to $\sim$41~kpc at the adopted distance. The difference in the line-of-sight velocities is $\Delta V_{los}$ $\sim$ 81~km\,s$^{-1}$. This suggests that M-8546-12704 is
not an isolated galaxy but a member of a galaxy pair.  

\begin{figure}
\resizebox{1.00\hsize}{!}{\includegraphics[angle=000]{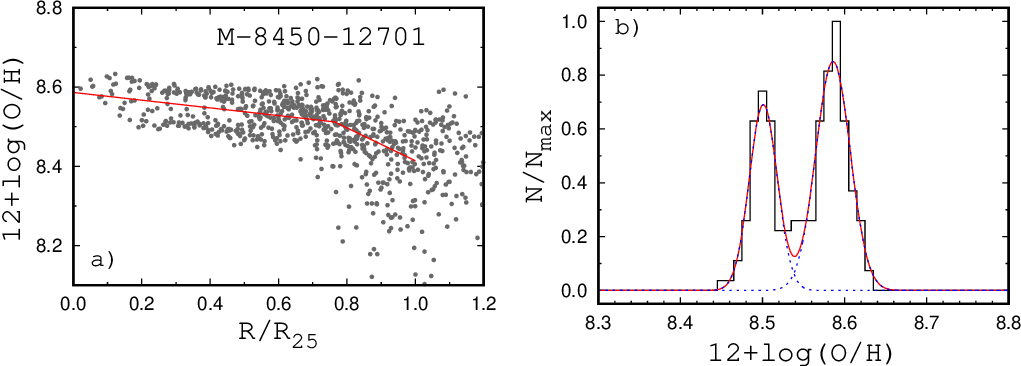}}
\caption{Oxygen abundance in galaxy M-8450-12701.
  {\em Panel} {\bf a:} Oxygen abundance in individual spaxel determined using R calibration from \citet{Pilyugin2016} as a function of radius (points). The line shows the broken
  linear relation for the radial abundance distribution within optical radius.
  {\sl Panel} {\bf b:} Normalised histogram of oxygen abundances for spaxels within the region of distinct azimuthal asymmetry ($<$ 0.5~$R_{25}$ for this galaxy).  The solid line denotes
  the histogram of the obtained oxygen abundances. The dashed red curve corresponds to the fit by two Gaussians to those data, the blue dotted curves show individual Gaussians.
}
\label{figure:m-8450-12701-oh}
\end{figure}

Panel (a) of Fig.~\ref{figure:m-8450-12701-oh} shows the oxygen abundance in the spaxel as a function of radius for M-8450-12701. The radial distribution of the oxygen abundances is usually
fitted by the linear relation of the type 12 +log(O/H) = (O/H)$_{0}$ + grad $\times$ $R/R_{25}$,  where (O/H)$_{0}$ = 12 + log(O/H)$_{\rm R=0}$ is the central oxygen abundance, and the gradient
is expressed in units of dex~$R_{25}^{-1}$.  The radial abundance distribution in M-8450-12701 should be fitted by the broken linear relation with the break at $R/R_{25}$ = 0.77. The mean
deviation of the points from the O/H -- $R$ relation is around 0.05 dex. The points with the deviations larger than 0.15 dex (3$\sigma$) are not used in the determination of the final
O/H -- $R$ relation. The parameters  (O/H)$_{0}$ and grad for the relations inside and outside break radius are reported in Table~\ref{table:general}. The obtained (O/H) -- $R$ relation is
shown by the line in the panel (a) of Fig.~\ref{figure:m-8450-12701-oh}. 

\begin{figure}
\resizebox{1.00\hsize}{!}{\includegraphics[angle=000]{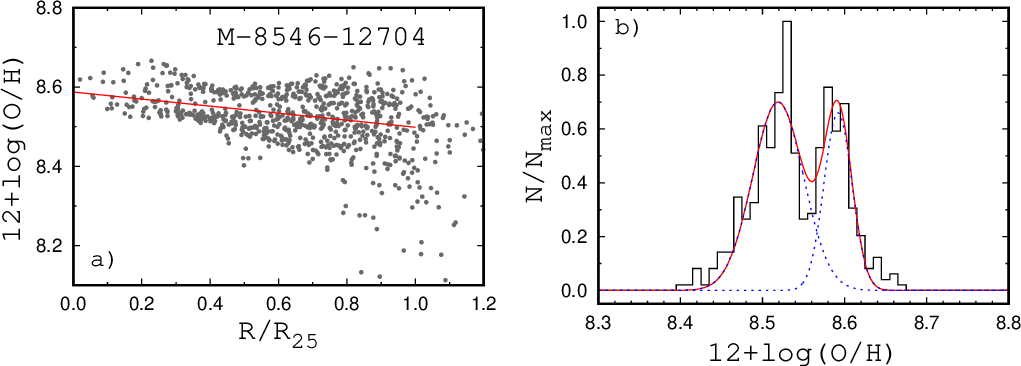}}
\caption{Same as Fig.~\ref{figure:m-8450-12701-oh}, but for galaxy M-8546-12704.
}
\label{figure:m-8546-12704-oh}
\end{figure}

\begin{figure}
\resizebox{1.00\hsize}{!}{\includegraphics[angle=000]{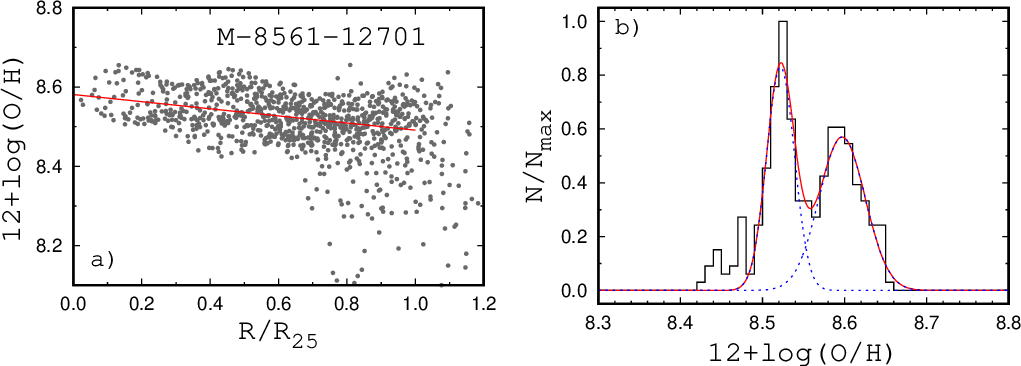}}
\caption{Same as Fig.~\ref{figure:m-8450-12701-oh}, but for galaxy M-8561-12701.
}
\label{figure:m-8561-12701-oh}
\end{figure}

\begin{figure}
\resizebox{1.00\hsize}{!}{\includegraphics[angle=000]{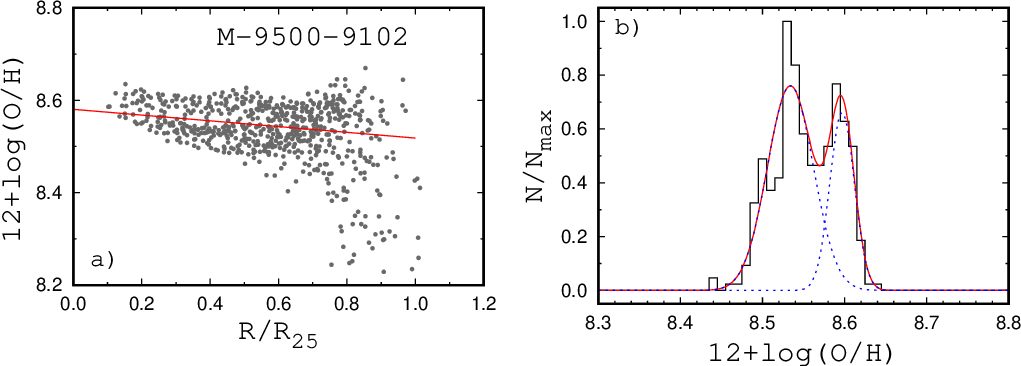}}
\caption{Same as Fig.~\ref{figure:m-8450-12701-oh}, but for galaxy M-9500-9102.
}
\label{figure:m-9500-09102-oh}
\end{figure}

\begin{figure}
\resizebox{1.00\hsize}{!}{\includegraphics[angle=000]{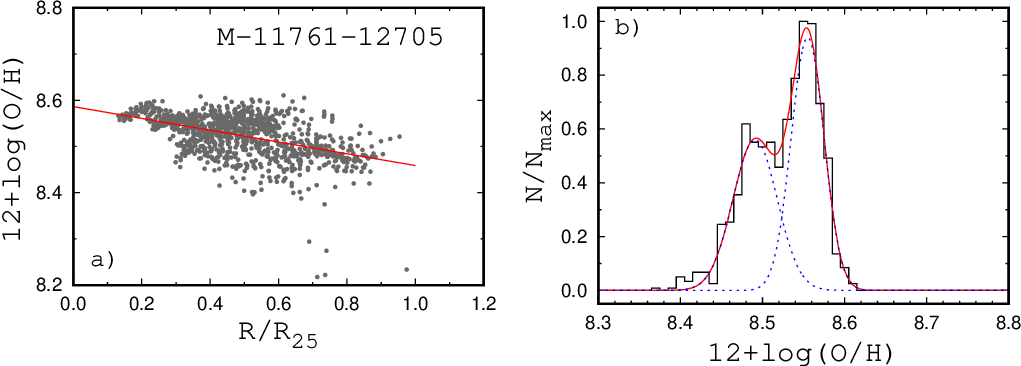}}
\caption{Same as Fig.~\ref{figure:m-8450-12701-oh}, but for galaxy M-11761-12705.
}
\label{figure:m-11761-12705-oh}
\end{figure}

One can see in panel (a) of Fig.~\ref{figure:m-8450-12701-oh} that the oxygen abundances in the inner part of the galaxy ($R \la 0.5R_{25}$) do not form a single sequence along the radius;
instead, they are split into two branches. Abundances of both the upper branch, (O/H)$_{\rm h}$, and the lower branch, (O/H)$_{\rm l,}$ are concentrated on two levels of the abundance.  
Panel (b) of Fig.~\ref{figure:m-8450-12701-oh} shows the normalised histogram of the oxygen abundances for the spaxels within region of azimuthal O/H asymmetry ($<$ 0.5$R_{25}$). 
The solid line denotes the histogram of the obtained oxygen abundances. The histogram demonstrates two distinct peaks in the abundance distribution for the spaxels in the inner
part of the galaxy. The Pearson's $\chi^{2}$ test shows that the null hypothesis can be rejected at the level of p $<$ 0.01. Fitting this distribution by two Gaussians (the dashed red 
line in panel (c) of Fig.~\ref{figure:m-8450-12701-oh}), we found the peak of high abundances (upper branch) is at 12+log(O/H)$_{\rm h}$ = 8.586, and the peak of the low abundances
(lower branch) is at 12+log(O/H)$_{\rm l}$ = 8.501.

Figures~\ref{figure:m-8546-12704-oh}-\ref{figure:m-11761-12705-oh} show the oxygen abundance as a function of radius and the normalised histogram of the oxygen abundances for the spaxels
in the region of azimuthal O/H asymmetry (radii of regions were reported above) for other target galaxies: M-8546-12704 (Fig.~\ref{figure:m-8546-12704-oh}), M-8546-12704
(Fig.~\ref{figure:m-8561-12701-oh}),  M-9500-09102 (Fig.~\ref{figure:m-9500-09102-oh}), and  M-11761-12705 (Fig.~\ref{figure:m-11761-12705-oh}). The radial abundance distribution in each
galaxy is approximated by the single linear relation. The histograms demonstrate two distinct peaks in the abundance distribution for each target galaxy. The Pearson's $\chi^{2}$ test
shows that the null hypothesis can be rejected at the level of p $<$ 0.01 for each target galaxy. The obtained characteristics of these galaxies (parameters (O/H)$_{0}$ and grad for the
O/H -- $R$ relation,  the peak values of the high (O/H)$_{\rm h}$ and the low (O/H)$_{\rm l}$ abundances in the normalised histogram)  are listed in Table~\ref{table:general}. 

\begin{figure}
\resizebox{1.00\hsize}{!}{\includegraphics[angle=000]{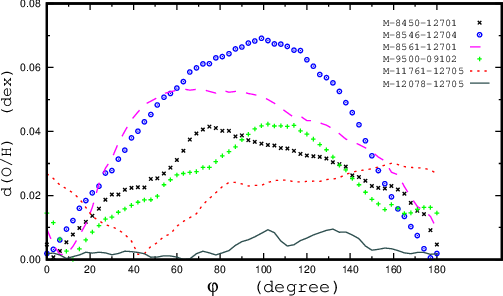}}
\caption{Global azimuthal asymmetry in oxygen abundance distribution across the whole disc of each target galaxy and control galaxy (M-12078-12705). The lines and symbols show the absolute
  value of difference between the arithmetic means of the deviations from the O/H -- $R$ relation for the spaxels within the sector, with azimuthal angles from $\varphi$ to $\varphi$ + 180
  and from  $\varphi$ + 180 to  $\varphi$ + 360 as a function of angle  $\varphi$.
}
\label{figure:oh-asym}
\end{figure}

We also determined the global azimuthal asymmetry in the oxygen abundance distribution across the whole disc of each target  galaxy and the control galaxy following \citet{Zinchenko2016}.
We divided a galaxy into two semi-circles by a dividing line at a position angle $\varphi$. For a fixed value of the angle $\varphi$, we determined the arithmetic mean of the deviations
d(O/H)$_{1}$ from the O/H -- $R$ relation for spaxels with azimuthal angles from $\varphi$ to $\varphi+180$ (first semi-circle) and the mean deviation d(O/H)$_{2}$ for spaxels with azimuthal
angles from $\varphi+180$ to $\varphi+360$ (second semi-circle). The absolute value of the difference d(O/H)$_{\rm asym}$ = d(O/H)$_{1}$ - d(O/H)$_{2}$ was estimated for different values of $\varphi$
(see Fig.~\ref{figure:oh-asym}). The position of $\varphi$ was counted anticlockwise from the major kinematic axis and changed with a step size of 3$\degr$ in the computations.
The maximum absolute value of the difference d(O/H)$_{\rm asym}$ was used to specify the global azimuthal asymmetry in the abundance distribution across the whole galaxy.
The obtained values of d(O/H)$_{\rm asym}$ are reported in Table~\ref{table:general}. The value of the d(O/H)$_{\rm asym}$ for the control galaxy is below 0.01, Fig.~\ref{figure:oh-asym}.
This is in line with the conclusion in \citet{Zinchenko2016} that the values of azimuthal asymmetry in galaxies are usually within 0.02. The values of the d(O/H)$_{\rm asym}$ for the target
galaxies range from $\sim$0.03 to $\sim$0.07 and are higher than in the control galaxy. However, the value of the global azimuthal O/H asymmetry for the whole galaxy d(O/H)$_{\rm asym}$
is lower than the difference between the high and low abundances in the region of distinct azimuthal O/H asymmetry d(O/H)$_{\rm h,l}$ = (O/H)$_{\rm h}$ -- (O/H)$_{\rm l}$. 
This is not surprising since the values of the  d(O/H)$_{\rm asym}$ and the d(O/H)$_{\rm h,l}$ should be close to each other if only the area of enhanced oxygen abundances coincides with the
semi-circle; this is not the case for target galaxies (see Fig.~\ref{figure:oh-map}). 

\section{Discussion}

\subsection{Galaxies with azimuthal abundance asymmetry among other galaxies}

\begin{figure}
\resizebox{1.00\hsize}{!}{\includegraphics[angle=000]{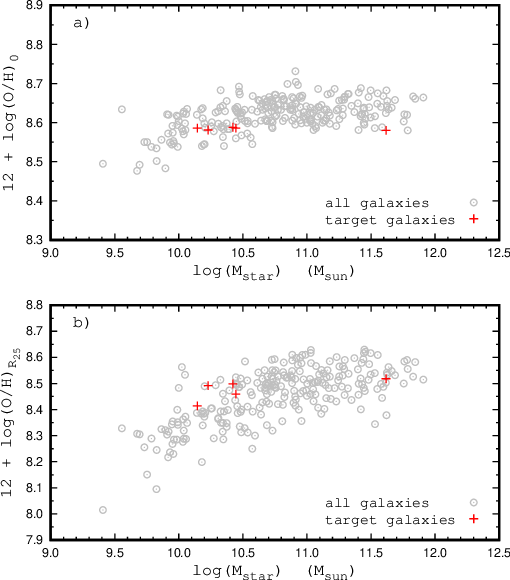}}
\caption{Comparison of oxygen abundances in galaxies with global azimuthal asymmetry with that in other galaxies.
  The central oxygen abundance (O/H)$_{0}$ ({\em panel} {\bf a}) and the oxygen abundance at the optical radius (O/H)$_{R_{25}}$ ({\em panel} {\bf b}) as a function of the
  stellar mass are shown. The grey points stand for the comparison galaxies. 
  The plus signs show the galaxies with an azimuthal abundance asymmetry. 
}
\label{figure:msp-oho-oh25}
\end{figure}

\begin{figure}
\resizebox{1.00\hsize}{!}{\includegraphics[angle=000]{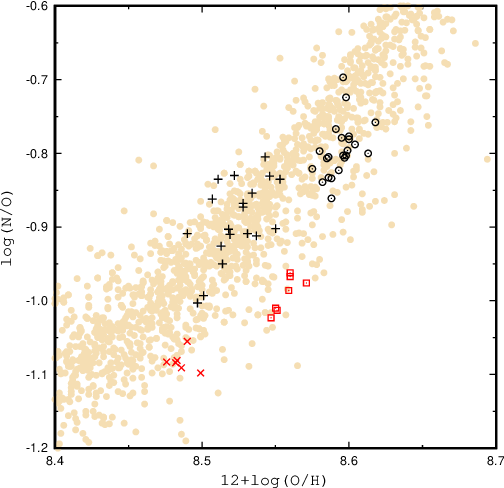}}
\caption{
  Locations of high- and low-metallicity regions of galaxies with azimuthal abundance asymmetry in the N/O -- O/H diagram.
  The circles designate the (O/H)$_{\rm h}$ regions in the moderate-mass galaxies, and   the plus signs show the (O/H)$_{\rm l}$ regions in these galaxies.
  The squares denote the (O/H)$_{\rm h}$ regions in the massive galaxy M-11761-12705, and the crosses mark the (O/H)$_{\rm l}$ regions in this galaxy. 
  The yellow points show H\,{\sc ii} regions in nearby galaxies (compilation in \citet{Pilyugin2016}). 
}
\label{figure:oh-no}
\end{figure}

Here, we compare our sample of galaxies with azimuthal abundance asymmetry to other galaxies. Firstly, we examined the evolutionary status of our galaxies; that is, we considered
their positions in the stellar mass versus star formation rate diagram.  We estimated the global star formation rate from the H$\alpha$ luminosity of a galaxy $L_{{\rm H}{\alpha}}$
using the calibration relation of \citet{Kennicutt1998} reduced by \citet{Brinchmann2004} for the Kroupa initial mass function \citep{Kroupa2001}.  
The  H$\alpha$ luminosity of a galaxy $L_{{\rm H}{\alpha}}$ was determined as a sum of the H$\alpha$ luminosities of the spaxels with  H\,{\sc ii}-region-like spectra within the optical
radius.  \citet{Speagle2014} investigated the evolution of the star-forming galaxy main sequence in the SFR -- $M_{\star}$ diagram using a compilation of 25 studies from the
literature. They found the `consensus' relation  SFR = $f$($M_{\star},t$). The SFRs in galaxies with azimuthal abundance asymmetry agree with the SFR -- $M_{\star}$ relation for
the present-day epoch ($t$ = 13.6 Gyr).

Next, we compare the central oxygen abundance (O/H)$_{0}$ and the oxygen abundance at the optical radius (O/H)$_{R_{25}}$ abundances in the galaxies with azimuthal abundance asymmetry
and in other galaxies, Fig.~\ref{figure:msp-oho-oh25}. The oxygen abundance in the high-metallicity region in the galaxies with azimuthal O/H asymmetry, (O/H)$_{\rm h}$, is close
to central oxygen abundance.  Inspection of the upper panel of Fig.~\ref{figure:msp-oho-oh25} shows that the (O/H)$_{0}$ in the massive galaxy M-11761-12705 is lower than the central
oxygen abundances in galaxies of similar masses, while the (O/H)$_{0}$ in four moderate-mass galaxies is comparable to the (O/H)$_{0}$ in galaxies of similar masses. This indicates
that the origin of the high-metallicity regions in the massive galaxy M-11761-12705 and four moderate-mass galaxies can be different. An examination of the lower panel of
Fig.~\ref{figure:msp-oho-oh25} shows that the (O/H)$_{R_{25}}$ in the massive galaxy M-11761-12705 is located in the lower envelope of the band in the $M_{\star}$ -- (O/H)$_{R_{25}}$ diagram,
while the (O/H)$_{l}$ of four moderate-mass galaxies is located in the upper envelope of the band. 

The N/O value for a given O/H contains important information about the enrichment history of heavy elements in the region.  The N/O -- O/H (or N/H--O/H) diagram has been a subject of
many investigations \citep[][among many others]{Edmunds1978,Pilyugin1992,Pilyugin1993,Gavilan2006,Pilyugin2011,Vincenzo2016,Berg2019,Schaefer2020,Schaefer2022,Roy2021,Johnson2023}.
Since the nitrogen production is secondary at high metallicities (12 + log(O/H) $\ga$ 8.0) then the nitrogen-to-oxygen ratio increases with metallicity at high oxygen abundances. 
Fig.~\ref{figure:oh-no} shows the positions of the high- and low-metallicity regions of our sample of galaxies in the O/H -- N/O diagram.  The position of the region in the O/H -- N/O
diagram is mainly defined by three factors. First, the N/O value at a given O/H is dependent on the star formation history \citep[e.g.][]{Edmunds1978,Pilyugin1992,Pilyugin2011,Maiolino2019,Romano2022}.
Since oxygen and nitrogen are produced in stars of different masses, there is a significant time delay between the release of oxygen, which is mainly produced in massive stars, and
that of nitrogen, which is produced in intermediate- and low-mass stars, into the interstellar medium. The N/O ratio of a region is an indicator of the time that has elapsed since
the last episode of star formation. When the bulk of the nitrogen-producing stars from previous generations complete their evolution, the nitrogen abundance reaches a high level and the
region is located in the upper envelope of the O/H -- N/O diagram.

Second, galactic winds alter the N/O abundance ratio in the region \citep[e.g.][]{Pilyugin1993,Vincenzo2016} and other characteristics of galaxies  \citep[e.g.][]{Dekel1986,MacLow1999,Sharda2021}.
The galactic wind may consist of two parts: the ordinary galactic wind and the enriched galactic wind. Galactic winds are caused by the collective effect of supernova explosions.
The ordinary galactic wind is the ambient interstellar medium leaving the galaxy. Some part of supernova ejecta may also leave the galaxy as a part of the galactic wind that drove them.
This part of supernova ejections can be called the enriched galactic wind because the supernova ejecta are enriched in heavy elements. The enriched galactic winds result in  an increase
of the N/O abundance ratio because the supernova ejecta are the main source of the enrichment of the interstellar medium in oxygen, while they do not make a significant contribution to
the nitrogen enrichment. The ordinary galactic winds also result  in an increase in N/O abundance ratio, because the nitrogen ejected by the long-lived intermediate- and low-mass stars
of previous star generations is mixed with the smaller amount of interstellar gas compared to oxygen ejected by massive stars of those generations. One can say that the fraction of oxygen
produced by previous star generations leaving the galaxy through the ordinary galactic winds is higher than the fraction of nitrogen produced by those star generations.     

Third, the infall of low-metallicity (pristine) gas into the region changes its position in the N/O -- O/H diagram \citep{Koppen2005, Maiolino2019}. Indeed, the pristine gas mixed to the
ambient interstellar gas reduces the absolute values of the nitrogen and oxygen abundances, but it does not change the N/O ratio. 

An inspection of Fig.~\ref{figure:oh-no} shows that the positions of the high- and low-metallicity regions in the massive galaxy M-11761-12705 and moderate-mass galaxies  in the
O/H -- N/O diagram are significantly different. Since the star formation history and efficiency of galactic winds are correlated with the galactic mass, the locations of galaxies
of different masses in the O/H -- N/O diagram are different, with higher-mass systems having higher N/O at a fixed O/H \citep{PerezMontero2013,Schaefer2022}. However, the position of
the massive galaxy M-11761-12705 with azimuthal abundance asymmetry  does not follow this trend.
 
\subsection{On the origin of the azimuthal abundance asymmetry in the massive galaxy M-11761-12705}

\begin{figure}
\resizebox{1.00\hsize}{!}{\includegraphics[angle=000]{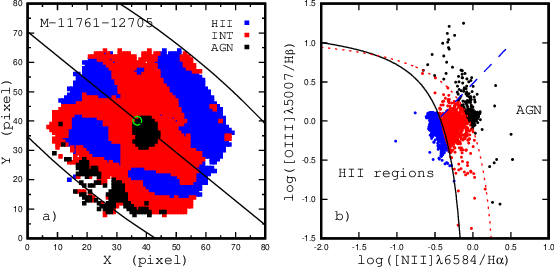}}
\caption{
BPT types of spaxel spectra of MaNGA galaxy M-11761-12705.
{\em Panel} {\bf a:} Map of BPT types of spaxel spectra. The BPT radiation types for individual spaxels are colour-coded. The circle shows the kinematic
centre of the galaxy, the line indicates a position of the major kinematic axis of the galaxy, and the ellipse is the optical radius. 
{\em Panel} {\bf b:}  BPT diagram for individual spaxels with H\,{\sc ii}-region-like (blue), intermediate (red), and AGN-like (black) spectrum classification.
The solid and short-dashed curves mark the demarcation line between AGNs and H\,{\sc ii} regions defined by \citet{Kauffmann2003}
and \citet{Kewley2001}, respectively. The long-dashed line is the dividing line between Seyfert galaxies and LINERs defined by \citet{CidFernandes2010}.
}
\label{figure:m-11761-12705-bpt}
\end{figure}

It was noted above that there is evidence that the origin of the high- and low-metallicity regions in the massive galaxy M-11761-12705 and in moderate-mass galaxies can be different.
Here, we consider the giant galaxy M-11761-12705, which has a stellar mass of log($M_{\star}/M_{\sun}$) = 11.6. The field of view of the MaNGA observation only covers a part of M-11761-12705.
A large fraction of the individual spaxels do not show the  H\,{\sc ii}-region-like spectra; in particular, the galaxy hosts the central AGN (see Fig.~\ref{figure:m-11761-12705-bpt}).
The oxygen abundances can be estimated in the spaxels with the  H\,{\sc ii}-region-like spectra. As a result, abundance map does not cover the whole  field of view of the MaNGA observation
(see panel (e) in Fig.~\ref{figure:oh-map}).

The central oxygen abundance in M-11761-12705 is lower than the central oxygen abundances in galaxies of similar masses. The high- and low-metallicity regions in M-11761-12705 are located
in the very low envelope of the band outlined by H\,{\sc ii} regions in nearby galaxies in the O/H -- N/O diagram in  Fig.~\ref{figure:oh-no}. This is just opposite the expected location,
since massive galaxies are usually well advanced in their (chemical) evolution and occupy the upper envelope of the band in the O/H -- N/O diagram \citep{PerezMontero2013,Schaefer2022,Pilyugin2024}. 

One can suggest a simple scenario for the origin of unusual abundances in M-11761-12705. As other massive galaxies, M-11761-12705  was well advanced in its (chemical)
evolution, converting a bulk of its gas into stars. The low-metallicity gas infall took place onto the galaxy, decreasing the abundance in the ambient interstellar gas. The subsequent
starburst occurred some time ago, between 20~Myr $\la$ $t$ $\la$ 50~Myr. This time is sufficient for the massive stars to end their evolution as supernova explosions and to enrich 
the interstellar medium with oxygen. At the same time, a bulk of the nitrogen-producing stars do not have enough time to complete their evolution and eject nitrogen into the interstellar
medium \citep[see e.g. Fig. 1 in][]{Maiolino2019}. As a result, the N/O ratio becomes very low. The gas infall onto M-11761-12705 and starburst can be caused by the interaction with
another galaxy. Indeed, M-11761-12705 is a member of the galaxy pair \citep{Tempel2018}. The interaction is also confirmed by the high value of the asymmetry parameter, $A$, (see below). 
Our scenario is similar to that considered by \citet{PerezDiaz2024}. They found that massive luminous infrared galaxies are located in the low envelope of the band in the O/H -- N/O diagram.
\citet{PerezDiaz2024} concluded that their positions can be caused by an infall of metal-poor gas eventually followed by a rapid enrichment.
 
\subsection{Origin of the azimuthal abundance asymmetry in moderate-mass galaxies of our sample}

\begin{figure*}
\resizebox{1.00\hsize}{!}{\includegraphics[angle=000]{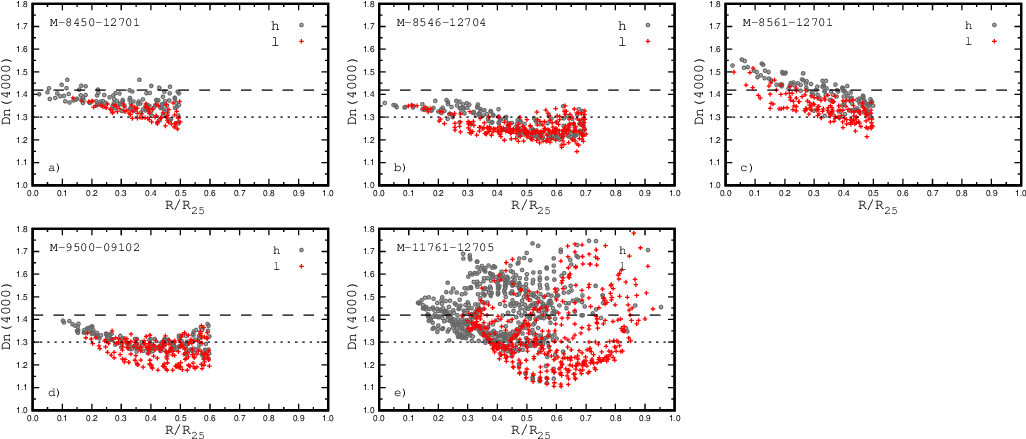}}
\caption{
  Spectral index D$_{n}$(4000) for individual spaxel as a function of radius for our sample of galaxies.
  The grey circles denote the spaxels of the high-metallicity regions,  the red plus signs are  the spaxels of the low-metallicity regions.
  The dashed line marks D$_{n}$(4000) = 1.42, and the dotted line shows D$_{n}$(4000) = 1.3 (see text).  
}
\label{figure:rg-dn}
\end{figure*}

\begin{figure*}
\resizebox{1.00\hsize}{!}{\includegraphics[angle=000]{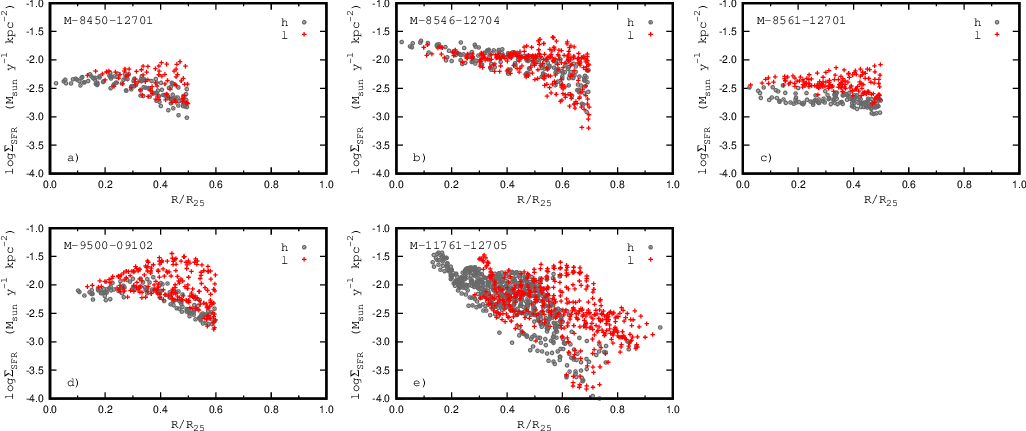}}
\caption{
  Star formation surface density for individual spaxel as a function of radius for our sample of galaxies.
  The grey circles denote the spaxels of the high-metallicity regions,  and the red plus signs show  the spaxels of the low-metallicity regions.
}
\label{figure:rg-sfr}
\end{figure*}

The stellar masses of four galaxies in our sample of galaxies with azimuthal abundance asymmetry lie in the range from log($M_{\star}/M_{\sun}$) = 10.1 (galaxy M-8450-12701) to
log($M_{\star}/M_{\sun}$) = 10.4 (galaxy M-8561-12701). Here, we analyse properties of those galaxies in order to understand  the origin of the azimuthal asymmetry in abundance.  

In each galaxy, the radial abundance distributions within the high- and low-metallicity regions are flat (the abundances are at approximately constant levels). 
The positions of high-metallicity regions of the galaxies M-8450-12701, M-8561-12701, and M-9500-09102 in the N/O -- O/H diagram are closer to the lower envelope of the band
outlined by the  H\,{\sc ii} regions in nearby galaxies than the positions of low-metallicity regions of those galaxies, Fig.~\ref{figure:oh-no}. 
This effect is less evident for the galaxy M-8546-12704. 

The simultaneous enhancement of the oxygen abundance in high-metallicity region and shift of its position towards the low envelope in the N/O -- O/H diagram in comparison to
the low-metallicity region cannot be explained by the continuous star formation; instead, those facts together suggest that the enrichment in oxygen and nitrogen should be
attributed to the starburst in the high-metallicity region. If the starburst occurred several dozens Myr ago, a large number of massive stars ended their evolution and enriched
the interstellar medium with oxygen (producing a jump-like enhancement of the oxygen abundance), while many nitrogen-producing stars do not have enough time to complete their evolution
and eject nitrogen into the interstellar medium (resulting in the shift towards the lower envelope in the N/O -- O/H diagram). The difference in oxygen abundances between the
high-metallicity and low-metallicity regions of 0.08~dex corresponds to the change in the gas-mass fraction (astration level) of $\sim$0.06. The amount of stellar mass formed
during the starburst can range from several to several dozens percentage \citep{French2018,Reeves2024}.

Thus, the origin of the bimodality of oxygen abundances in galaxies with azimuthal asymmetry can be the following. The oxygen and nitrogen abundances were uniform (similar to
those in the low-metallicity region) in the whole galaxy. An episode of high star formation (starburst) occurred several dozens of Myr ago in a large region, increasing the
oxygen abundance and (temporarily) decreasing the nitrogen-to-oxygen ratio in this region.

The spectral index D$_{n}$(4000) (the ratio of the average flux density in the bands 3850-3950 and 4000-4100 \AA) can serve as a proxy for the look-back time of the starburst
\citep[e.g.][]{French2018,Wu2023}. \citet{French2018} noted that the relation between D$_{n}$(4000) and the post-burst age suffers from a degeneracy with the burst mass fraction
and burst duration. They find that for post-starbursts with D$_{n}$(4000) $<$ 1.3, where post-starburst ages are typically under 300~Myr, the degeneracy is lessened, and D$_{n}$(4000)
is highly correlated with the post-burst age. However, if D$_{n}$(4000) $>1.42$, where post-starburst ages typically range from 300 to 1500~Myr, D$_{n}$(4000) is no longer significantly
correlated with the post-burst age.

Figure~\ref{figure:rg-dn} shows the spectral index D$_{n}$(4000) for individual spaxels as a function of the radius for our sample of galaxies. The grey circles denote the spaxels of
high-metallicity regions,  the red plus signs are the spaxels of low-metallicity regions. An inspection of Fig.~\ref{figure:rg-dn} shows that the spectral index D$_{n}$(4000) is higher
in the spaxels of high-metallicity regions than in the spaxels of low-metallicity regions; that is, the ages of the stellar populations are higher in the spaxels of high-metallicity
regions than in the spaxels of low-metallicity regions. The galaxy M-8561-12701 shows  most clearly that the spectral index D$_{n}$(4000) is higher in the spaxels of high-metallicity
regions than in those of low-metallicity regions (see panel (c) of Fig.~\ref{figure:rg-dn}). This is in conflict with the assumption that the enhancement of oxygen abundance in the
high-metallicity region should be attributed to the starburst that occurred several dozens Myr ago. Fig.~\ref{figure:rg-sfr} shows the current star formation rate surface density for
individual spaxels as a function of the radius for our sample of galaxies. Fig.~\ref{figure:rg-sfr}, panel (c), shows that the current star formation rate is higher in the spaxels of
low-metallicity regions than in those of high-metallicity regions.  Again, the galaxy M-8561-12701 shows this difference most clearly. A comparison between panel (c) of Fig.~\ref{figure:rg-dn}
and panel (c) of Fig.~\ref{figure:rg-sfr} shows that the spectral indices of D$_{n}$(4000) are lower in spaxels with the high current star formation rate than in spaxels with the
low current star formation rate. This can be considered as evidence that the spectral index D$_{n}$(4000) for the individual spaxel can be significantly affected by current star
formation and is not a very reliable indicator of stellar age. 

If the assumption about the starburst in the high-metallicity region several dozens Myr ago should be rejected based on the behaviour of the  D$_{n}$(4000) indices, the higher oxygen
abundance in the high-metallicity region together with the shift of its position towards the low envelope in the N/O -- O/H diagram compared to the low-metallicity region should be
explained in an alternative way: the star formation in the galaxy is accompanied by galactic winds, and the region that evolved with the higher efficiency of the enriched galactic
winds shows lower metallicity. It was discussed above that the enriched galactic winds decrease the O/H abundance and increase the N/O abundance ratio in the region.  Galactic winds
are strongly dependent on M$_{\star}$, star formation rate, and characteristics of the interstellar medium \citep{Hayward2017,Marasco2023}. In the contemporary Universe, strong winds
are only observed in galaxies undergoing intense bursts of star formation \citep{Heckman2015}. There is a star formation rate threshold for the generation of galactic winds,
$\Sigma_{\rm SFR}$ $\sim$ 0.1$M_{\sun}$ y$^{-1}$ kpc$^{-2}$ \citep{Heckman2001} or $\Sigma_{\rm SFR}$ $\sim$ 0.01$M_{\sun}$ y$^{-1}$ kpc$^{-2}$ \citep{LopezCoba2019}. It should be noted that
$\Sigma_{\rm SFR}$ $\sim$ 0.1$M_{\sun}$ y$^{-1}$ kpc$^{-2}$ is the dividing value of the SFR between the non-starbursting and starbursting galaxies \citep{Kennicutt2021}. 

Thus, the higher oxygen abundance in the high-metallicity region together with the shift in its position towards the low envelope in the N/O -- O/H diagram compared to the low-metallicity
region can be explained in two alternative ways. First,  the starburst in the high-metallicity region occurred several dozens Myr ago. Second, the star formation in the galaxy is
accompanied by the galactic winds, and the region evolved with the lower efficiency of the enriched galactic winds shows higher metallicity.   It should be noted that both scenarios suggest
episodes of high star formation (starburst). This suggests that the flat abundance distributions (abundances at a nearly constant level) in the regions of moderate-mass galaxies are related
to the bursty star formation. Of course, our sample of four galaxies is too small to draw a solid conclusion, and further investigations are necessary to confirm or reject this conclusion.

\subsection{Mergers and interactions}

\begin{table}
\caption[]{\label{table:a-ra}
  Asymmetry parameter, $A$, of a light distribution across the galaxy and index, $R_{A,}$  of the asymmetry in the distribution of the residual fluxes across the galaxy
  for the MaNGA galaxies of our sample. 
}
\begin{center}
\begin{tabular}{rcc} \hline \hline
MaNGA name        &
A                 &
R$_{A}$            \\
\hline
%  galaxy          A         AR       ARc    
  8450 12701  &  0.241  &  0.119    \\ 
  8546 12704  &  0.324  &  0.179    \\ 
  8561 12701  &  0.284  &  0.135    \\ 
  9500 09102  &  0.306  &  0.161    \\ 
 11761 12705  &  0.610  &  0.307    \\ 
                    \hline
\end{tabular}\\
\end{center}
\end{table}

Despite extensive observational and theoretical studies, the physical mechanisms that regulate the star formation rates of galaxies are still poorly understood \citep[e.g.][]{Yesuf2017}.
Therefore, an important question, that is, why the (chemical) evolutions of two sectors of the galaxy with azimuthal abundance asymmetry differ, cannot be answered. However, if there is
any other distinctive characteristic common to all these galaxies, it could be a hint as of the necessary condition for the origin of those galaxies. Currently, there is a belief that
galaxy-galaxy interactions and mergers can have a significant effect on galaxy evolution, altering different characteristics of a galaxy. In particular, the interaction and merger can
result in the enhancement of the star formation rate \citep[][among many others]{Barton2000,Lotz2008,Pawlik2016}. It is predicted that the galaxy-galaxy interactions and mergers can
produce a flat abundance gradient \citep{Rupke2010a,Sillero2017,Bustamante2018,Sharda2021}, and observations show that the merging and interacting systems exhibit shallow oxygen abundance
gradients compared to isolated spiral galaxies \citep{Rupke2010b,Kewley2010,Rosa2014,Croxall2015,TorresFlores2020}. Here, we examine whether our galaxies with azimuthal abundance asymmetry
show signs of interactions or mergers. It should be noted that we selected a sample of discy galaxies for which the curves of iso-velocities in the measured line-of-sight velocity fields
resemble a set of parabola-like curves (hourglass appearance of the rotation disc).  Using this criterion, we rejected strongly interacting and merging galaxies if the line-of-sight velocity
field was distorted to such an extent that the determination of the geometrical angles and rotation curve was impossible. Therefore, our sample can only include the interacting and merging
galaxies where the distortion of the line-of-sight velocity field is not crucial.   

The most common way to identify galaxy mergers and interactions is by morphology. The asymmetry parameter, $A$, quantifies the asymmetry of a light distribution across the galaxy
\citep{Schade1995,Conselice2003}: 
\begin{equation}
A = \frac{\sum\limits_{i,j} |I_{ij} - I_{ij}^{180}|}{\sum\limits_{i,j=1} I_{ij}}  
\label{equation:a}
,\end{equation}
where $I_{ij}$ is the flux of a spaxel in the original image, and $I_{ij}^{180}$ is the flux of the same spaxel after the image has been rotated by 180$\degr$ around the centre of the galaxy.
The sum is carried out on all spixels within the optical radius of the galaxy. We also estimated the indices, $R_{A,}$ to specify the distribution of residual fluxes \citep{Schade1995}: 
\begin{equation}
R_{A} = \frac{\sum\limits_{i,j} \frac{1}{2}|W_{ij} - W_{ij}^{180}|}{\sum\limits_{i,j=1} I_{ij}}  
\label{equation:ra}
,\end{equation}
where $W_{ij}$ is a residual spaxel flux\footnote{We used the symbol $W$ to designate the residual flux in the individual spaxel instead of the commonly used symbol $R,$ since it already used
  here to designate the galactocentric distance of the spaxel.} after the model flux is subtracted: $W_{ij}$ = $I_{ij}^{obs}$ --  $I_{ij}^{mod}$. The model flux in the spaxel was estimated from the
photometric profile using Eq.~(\ref{equation:decomp}). The determined value of parameters $A$ and $R_{A}$ for our sample of galaxies  are presented in Table~\ref{table:a-ra}.  

The average value of the asymmetry parameters in early-type spirals (Sa and Sb galaxies) is $A$ = 0.07$\pm$0.04; the late-type spirals (Sc and Sd galaxies) show higher asymmetries
of $A$ = 0.15$\pm$0.06 \citep{Conselice2003}. The high asymmetry value ($A >$ 0.35) can serve as an indicator of the merger or interaction  \citep{Conselice2003,Wilkinson2022}.  A galaxy
is classified as `asymmetric' if  $R_{A} > 0.05$  \citep{Schade1995}.

Two galaxies of our sample (M-8546-12704  and M-11761-12705) are members of galaxy pairs. However, the asymmetry parameter, $A$, is above the canonical threshold value ($A$ = 0.35) for
non-interacting/interacting galaxies in  the massive galaxy M-11761-12705 only (see Table~\ref{table:a-ra}).  The values of the $A$ parameter in four moderate-mass galaxies (including the
member of the galaxy pair, M-8546-12704) are higher than the average value of the asymmetry parameters in the late-type spirals, but they are below the threshold value. Thus, we cannot make
a solid conclusion as to whether the moderate-mass galaxies of our sample show signs of interactions or mergers and whether this is a necessary condition for the origin of those galaxies. 
At the same time, the values of the asymmetry index $R_{A,}$  in the distribution of the residual fluxes after the model flux is subtracted is higher than 0.05 in all the galaxies of our
sample; that is, each galaxy can be classified as asymmetric.  

Galaxy merger/interaction features may persist for up to $\sim$1~Gyr, but gradually fade and become faint $\sim$200~Myr after coalescence \citep{Lotz2008,Pawlik2016,Wilkinson2022}.
Recently, \citet{Bottrell2024} used the Illustris TNG50 simulation to investigate the relative roles of major mergers (stellar-mass ratios $\mu \ga$ 0.25), minor mergers (0.1
$\la \mu < $0.25), and mini mergers (0.01 $\la \mu <$ 0.1) in the evolution of star-forming galaxies. They found that individual mini-merger events yield small enhancements in asymmetries
($\Delta R_{A} \sim$0.01) that are sustained on long timescales ($\sim$3 Gyr after coalescence, on average). The remnants of major/minor mergers peak at much greater amplitudes
($\Delta R_{A} \sim$ 0.03-0.05), but this enhancement is short-lived (less than $\sim$1~Gyr after coalescence). Moreover, the timing of morphological disturbances is generally offset
from the peak in star formation rates, with strong morphological disturbances occurring before bursts of merger-induced star formation \citep{Lotz2008}. Hence, not all merger-induced
starbursts exhibit morphological features of the merger.

\section{Conclusion}

We find five MaNGA galaxies that show distinct azimuthal asymmetry in the abundance across the disc, in the sense that in the inner part (more than half of optical radius $R_{25}$)
of each galaxy there is a sector-like (up to semi-circle) region where the oxygen abundances (O/H)$_{\rm h}$ are higher than the abundances (O/H)$_{\rm l}$ in the other sector.
M-11761-12705 is a massive galaxy with a stellar mass of log($M_{\star}/M_{\sun}$) = 11.6. The masses of four other galaxies (M-8450-12701, M-8546-12704, M-8561-12701, M-9500-09102) are
moderate, within the range of 10.1 $\la$ log($M_{\star}/M_{\sun}$) $\la$ 10.4. 

Abundances within high- and low-metallicity regions show flat radial gradients (the abundances are at nearly constant levels in each region). The histogram for the spaxel
abundances demonstrates two distinct peaks, and the differences between the peaks are of 0.06 -- 0.08~dex. The high-metallicity regions are usually located in the O/H -- N/O diagram
closer to the lower envelope of the band outlined by H\,{\sc ii} regions in nearby galaxies than the low-metallicity regions. The abundance properties in the massive galaxy
M-11761-12705 can be explained by the low-metallicity gas infall onto the galaxy and subsequent episode of high star formation rate (starburst) that occurred between 20~Myr
$\la$ $t$ $\la$ 50~Myr ago in the diluted interstellar medium.  

For moderate-mass galaxies, the higher oxygen abundance in the high-metallicity region and its shift towards the lower envelope of the band in the N/O -- O/H diagram in comparison to
the low-metallicity region can be explained in one of two ways: either the starburst in the high-metallicity region occurred several dozens of Myr ago; or, the star formation in
the galaxy is accompanied by the galactic winds, and  the region evolved with the lower efficiency of the enriched galactic winds shows higher metallicity. Both scenarios suggest an
episode with a high star formation rate (starburst). This suggests that the flat abundance distributions in the regions of moderate-mass galaxies are related to the bursty star
formation. However, our sample of four galaxies is too small to make a solid conclusion; further investigations are necessary.

Two galaxies of our sample are members of galaxy pairs. However, the asymmetry parameter, $A$, quantifying asymmetry of a light distribution across the galaxy, is above the canonical
threshold value ($A$ = 0.35) for non-interacting/interacting galaxies in the massive galaxy M-11761-12705 only.  At the same time, the values of the index $R_{A}$ of the asymmetry
in the distribution of the residual fluxes after a model flux is subtracted are higher than 0.05 in all the galaxies of our sample; that is, each galaxy can be classified as asymmetric.  

\begin{acknowledgements}

We are grateful to the referee for his/her constructive comments. \\ 
L.S.P acknowledges support from the Research Council of Lithuania (LMTLT, No. P-LU-PAR-25-8). \\
This research has made use of the NASA/IPAC Extragalactic Database (NED), which
is funded by the National Aeronautics and Space Administration and operated by
the California Institute of Technology.  \\
We acknowledge the usage of the HyperLeda database (http://leda.univ-lyon1.fr). \\
Funding for SDSS-III has been provided by the Alfred P. Sloan Foundation,
the Participating Institutions, the National Science Foundation,
and the U.S. Department of Energy Office of Science.
The SDSS-III web site is http://www.sdss3.org/. 
Funding for the Sloan Digital Sky Survey IV has been provided by the
Alfred P. Sloan Foundation, the U.S. Department of Energy Office of Science,
and the Participating Institutions. SDSS-IV acknowledges
support and resources from the Center for High-Performance Computing at
the University of Utah. The SDSS web site is www.sdss.org. 
\end{acknowledgements}


\begin{thebibliography}{}

\bibitem [Abdurro'uf  et al. (2022)]{Abdurrouf2022}
          Abdurro'uf, Accetta  K., Aerts C., et al., 2022, ApJS, 259, 35

\bibitem [Baldwin et al.(1981)]{Baldwin1981}
          Baldwin J.A., Phillips M.M., \& Terlevich R. 1981, PASP, 93, 5
         
\bibitem [Barton et al.(2000)]{Barton2000} 
          Barton E.J., Geller M.J. Kenyon S.J., 2000, ApJ, 530. 660

\bibitem [Belfiore et al.(2017)]{Belfiore2017}
          Belfiore F., Maiolino R., Tremonti C., et al., 2017, MNRAS, 469, 151

\bibitem [Bellardini et al.(2021)]{Bellardini2021}
          Bellardini M.A., Wetzel A., Loebman S.R., et al., 2021, MNRAS, 505, 4586
  
\bibitem [Bellardini et al.(2022)]{Bellardini2022}
          Bellardini M.A., Wetzel A., Loebman S.R., Bailin J., 2022, MNRAS, 514, 4270 

\bibitem [Berg et al.(2019)]{Berg2019}
          Berg D.A., Erb D.K., Henry R.B.C., Skillman E.D., McQuinn K.B.W., ApJ, 874, 93
          
\bibitem [Berg et al.(2020)]{Berg2020}
          Berg D.A., Pogge R.W., Skillman E.D., et al.,  2020, ApJ, 893, 96 
          
\bibitem [Bottrell et al.(2024)]{Bottrell2024} 
          Bottrell C., Yesuf H.M., Popping G., et al., 2024, MNRAS, 527, 6506
  
\bibitem [Brinchmann et al. (2004)]{Brinchmann2004}
          Brinchmann J., Charlot S., White S.D.M., et al., 2004, MNRAS, 351, 1151 

\bibitem [Bruzual \& Charlot(2003)]{Bruzual2003} 
          Bruzual G., Charlot S., 2003, MNRAS, 344, 1000 
                      
\bibitem [Bundy et al.(2015)]{Bundy2015}
          Bundy K., Bershady M.A., Law D.R., et al., 2015, ApJ, 798, 7 
  
\bibitem [Bustamante et al.(2018)]{Bustamante2018} 
          Bustamante S., Sparre M., Springel V., Grand R.J.J., 2018, MNRAS, 479, 3381 

\bibitem [Cardelli et al. (1989)]{Cardelli1989}
          Cardelli J.A., Clayton G.C., Mathis J.S., 1989, ApJ, 345, 245 
             
\bibitem [Chen et al. (2012)]{Chen2012} 
          Chen Y-M., Kauffmann G., Tremonti C.A., et al., 2012, MNRAS, 421, 314
  
\bibitem [Cid Fernandes et al. (2010)]{CidFernandes2010}
          Cid Fernandes R.,  Stasi\'{n}ska G., Schlickmann M.S., et al., 2010, MNRAS, 403, 1036
   
\bibitem [Conselice(2003)]{Conselice2003} 
          Conselice C.J., 2003, ApJS, 147, 1

\bibitem [Croxall et al.(2015)]{Croxall2015} 
          Croxall K.V., Pogge R.W., Berg D.A., Skillman E.D., Moustakas J., 2015, ApJ, 808, 42
          
\bibitem [Dawson et al. (2013)]{Dawson2013} 
          Dawson K.~S., Schlegel D.~J., Ahn C.~P., et al., 2013, AJ, 145, 10 

\bibitem [Dekel \& Silk(1986)]{Dekel1986} 
          Dekel A., Silk J., 1986, ApJ, 303, 39 

\bibitem [Edmunds \& Pagel (1978)]{Edmunds1978}
          Edmunds M.G. \& Pagel B.E.J., 1978, MNRAS, 185, 77P

\bibitem [Epinat et al.(2008)]{Epinat2008}
          Epinat B., Amram P., Marcelin M., et al., 2008, MNRAS, 388, 500 
  
\bibitem [French et al.(2018)]{French2018} 
          French K.D., Yang Y., Zabludoff A., Tremonti C.A., 2018, ApJ, 862, 2 

\bibitem [Gavil{\'a}n et al.(2006)]{Gavilan2006}
          Gavil{\'a}n M., Moll{\'a} M., Buell J.F., 2006, A\&A, 450, 509

\bibitem [Grand et al.(2016)]{Grand2016} 
          Grand R.J.J., Springel V., Kawata D., et al., 2016. MNRAS, 460.  L94
         
\bibitem [Grasha et al.(2022)]{Grasha2022}
          Grasha K., Chen Q.H., Battisti A.J., et al., 2022, ApJ, 929, 118 
 
\bibitem [Gusev et al.(2012)]{Gusev2012} 
          Gusev A.S., Pilyugin L.S., Sakhibov F., et al., 2012, MNRAS, 424, 1930 
         
\bibitem [Hayward \& Hopkins(2017)]{Hayward2017} 
          Hayward C., Hopkins P.F., 2017, MNRAS, 465, 1682 

\bibitem [Heckman(2001)]{Heckman2001} 
          Heckman T.M., in ASP Conf. Ser. 240, Gas and Galaxy Evolution, ed. J.E. Hibbard, M Rupen, J.H. van Gorkom (San Francisco. CA: ASP), 345
         
\bibitem [Heckman et al.(2015)]{Heckman2015} 
          Heckman T.M., Alexandroff R.M., Borthakur S., Overzier R., Leitherer C., 2015, ApJ, 809, 147 
         
\bibitem [Ho et al.(2015)]{Ho2015}  
          Ho I.-T., Kudritzki R.-P., Kewley L.J., et al.,  2015, MNRAS, 448, 2030 

\bibitem [Ho et al.(2018)]{Ho2018}  
          Ho I.-T., Meidt S.E.,  Kudritzki R.-P., et al., 2018, A\&A, 618, A64 
  
\bibitem [Johnson et al.(2023)]{Johnson2023}  
          Johnson J.W., Weinberg D.H., Vincenzo F., Bird J.C., Griffith E.J., 2023, MNRAS, 520, 782

\bibitem [Kauffmann et al.(2003)]{Kauffmann2003}
          Kauffmann G., Heckman T.M., Tremonti C., et al., 2003b, MNRAS, 346, 1055
          
\bibitem [Kewley et al.(2001)]{Kewley2001}
          Kewley L.J., Dopita M.A., Sutherland R.S., Heisler C.A., Trevena J.  2001 ApJ, 556, 121
 
\bibitem [Kennicutt(1998)]{Kennicutt1998}
          Kennicutt R.C., 1998, ARA\&A, 36, 189
           
\bibitem [Kennicutt \& De Los Reyes(2021)]{Kennicutt2021}  
          Kennicutt R.C., De Los Reyes M.A.C., 2021, ApJ, 908, 61 

\bibitem [Kewley et al.(2010)]{Kewley2010} 
          Kewley L.J., Rupke D., Zahid H.J., Geller M.J., Barton E.J., 2010, ApJ, 721, L48 

\bibitem [K{\"o}ppen \& Hensler(2005)]{Koppen2005} 
          K{\"o}ppen J., Hensler G., 2005, A\&A, 434, 531 
         
\bibitem [Kreckel et al.(2019)]{Kreckel2019} 
          Kreckel K., Ho I.-T., Blanc G.A., et al., 2019, ApJ, 887, 80 

\bibitem [Kroupa(2001)]{Kroupa2001}
          Kroupa P., 2001, MNRAS, 322, 231

\bibitem [Levy et al.(2018)]{Levy2018} 
          Levy R.C., Bolatto A.D., Teuben P., 2018, ApJ, 860, 92
  
\bibitem [L{\'o}pez-Cob{\'a} et al.(2019)]{LopezCoba2019} 
          L{\'o}pez-Cob{\'a} C., S{\'a}nchez S.F., Bland-Hawthorn J., et al., 2019, MNRAS, 482, 4032 
          
\bibitem [Lotz et al.(2008)]{Lotz2008} 
          Lotz J.M., Jonsson P., Cox T.J., Primack J.R., 2008, MNRAS, 391, 1137
         
\bibitem [Mac Low \& Ferrara(1999)]{MacLow1999} 
          Mac Low  M.-M., Ferrara A., 1999, ApJ, 513, 142 

\bibitem [Maiolino \& Mannucci(2019)]{Maiolino2019} 
          Maiolino R., Mannucci F., 2019, A\&A Rev., 27. 3 
         
\bibitem [Makarov et al.(2014)]{Makarov2014} 
          Makarov D., Prugniel P., Terekhova N., Courtois H., Vauglin I., 2014, A\&A, 570, A13
         
\bibitem [Marasco et al.(2014)]{Marasco2023} 
          Marasco A., Belfiore F., Cresci G., et al., 2023, A\&A, 670, A92 

\bibitem [Orr et al.(2023)]{Orr2023}
          Orr M.E., Burkhart B., Wetzel A., et al., 2023, MNRAS, 521, 3708 
          
\bibitem [Pawlik et al.(2016)]{Pawlik2016} 
          Pawlik M.M., Wild V., Walcher C.J., et al., 2016, MNRAS, 456, 3032 

\bibitem [P{\'e}rez-D{\'i}az et al.(2024)]{PerezDiaz2024} 
          P{\'e}rez-D{\'i}az B.,  P{\'e}rez-Montero E., Fern{\'a}ndez-Ontiveros J.A., V{\'i}lchez J.M., Amor{\'i}n R., 2024, Nat. Astron., 8, 368

\bibitem [P{\'e}rez-Montero et al.(2013)]{PerezMontero2013} 
          P{\'e}rez-Montero E., Contini T., Lamareille F., et al., 2013, A\&A, 549, A25
  
\bibitem [Pilyugin (1992)]{Pilyugin1992} 
          Pilyugin L.S., 1992, A\&A, 260, 58 
          
\bibitem [Pilyugin (1993)]{Pilyugin1993} 
          Pilyugin L.S., 1993, A\&A, 277, 42 

\bibitem [Pilyugin et al.(2004)]{Pilyugin2004} 
          Pilyugin L.S., V\'{\i}lchez J.M., Contini T., 2004, A\&A, 425, 849 

\bibitem [Pilyugin et al.(2007)]{Pilyugin2007} 
          Pilyugin L.S., Thuan T.X., V\'{\i}lchez J.M., 2007, MNRAS, 376, 353 
          
\bibitem [Pilyugin \& Thuan (2011)]{Pilyugin2011} 
          Pilyugin L.S., Thuan T.X., 2011, ApJ, 726, L23
          
\bibitem [Pilyugin et al.(2014)]{Pilyugin2014} 
          Pilyugin L.S., Grebel E.K., Kniazev A.Y., 2014, AJ, 147, 131 

\bibitem [Pilyugin \& Grebel(2016)]{Pilyugin2016} 
          Pilyugin L.S., Grebel E.K., 2016, MNRAS, 457, 3678 
         
\bibitem [Pilyugin et al.(2018)]{Pilyugin2018} 
          Pilyugin L.S., Grebel E.K., Zinchenko I.A., et al.,  2018, A\&A, 613, A1 
          
\bibitem [Pilyugin et al.(2019)]{Pilyugin2019} 
          Pilyugin L.S., Grebel E.K., Zinchenko I.A., Nefedyev Y.A., V{\'\i}lchez J.M., 2019, A\&A, 623, A122 

\bibitem [Pilyugin et al.(2020)]{Pilyugin2020} 
          Pilyugin L.S., Grebel E.K., Zinchenko I.A., et al., 2020, A\&A, 639, A96 

\bibitem [Pilyugin et al.(2021)]{Pilyugin2021} 
          Pilyugin L.S., Cedr{\'e}s B., Zinchenko I.A., et al., 2021, A\&A, 653, A11 
                             
\bibitem [Pilyugin \& Tautvai\~{s}ien\.{e}(2024)]{Pilyugin2024} 
          Pilyugin L.S., Tautvai\~{s}ien\.{e} G., 2024, A\&A, 682, A41 
          
\bibitem [Reeves \& Hudson(2024)]{Reeves2024} 
          Reeves A,M.M., Hudson M.J., 2024, MNRAS, 527, 2037

\bibitem [Romano(2022)]{Romano2022} 
          Romano D., 2022, A\&A Rev., 30, 7 

\bibitem [Rosa et al.(2014)]{Rosa2014}   
          Rosa D.A., Dors O.L., Krabbe A.C., et al., 2014, MNRAS, 444, 2005 
  
\bibitem [Roy et al.(2021)]{Roy2021}   
          Roy A., Dopita M.A., Krumholz M.R., et al., 2021, MNRAS, 502, 4359 
          
\bibitem [Rupke et al.(2010a)]{Rupke2010a}       
          Rupke D.S.N., Kewley L.J., Barnes J.E., 2010a, ApJ, 710, 156
  
\bibitem [Rupke et al.(2010b)]{Rupke2010b}       
          Rupke D.S.N., Kewley L.J., Chien L.-H., 2010b, ApJ, 723, 1255 
          
\bibitem [Sakhibov et al.(2018)]{Sakhibov2018}       
          Sakhibov F., Zinchenko I.A., Pilyugin L.S., et al., 2018, MNRAS, 474, 1657 
         
\bibitem [S\'{a}nchez et al.(2012)]{Sanchez2012} 
          S\'{a}nchez S.F., Kennicutt R.C., Gil de Paz A., et al.,  2012, A\&A, 538, A8

\bibitem [S\'{a}nchez et al.(2014)]{Sanchez2014}  
          S\'{a}nchez S.F., Rosales-Ortega F.F., Iglesias-P\'{a}ramo J., et al.,  2014, A\&A, 563, 49 
  
\bibitem [S{\'a}nchez et al.(2015)]{Sanchez2015}       
          S{\'a}nchez S.F., Galbany L., P{\'e}rez E., et al., 2015, A\&A, 573, A105 
        
\bibitem [S\'{a}nchez-Menguiano et al.(2016)]{SanchezMenguiano2016}  
          S{\'a}nchez-Menguiano L., S{\'a}nchez S.F., P{\'e}rez I., et al., 2016, A\&A, 587, A70
          
\bibitem [S{\'a}nchez-Menguiano et al.(2017)]{SanchezMenguiano2017}       
          S{\'a}nchez-Menguiano L., S{\'a}nchez S.F., P{\'e}rez I., et al., 2017, A\&A, 603, A113 
              
\bibitem [S\'{a}nchez-Menguiano et al.(2018)]{SanchezMenguiano2018}  
          S{\'a}nchez-Menguiano L., S{\'a}nchez S.F., P{\'e}rez I., et al., 2018, A\&A, 609, A119 
          
\bibitem [S{\'a}nchez-Menguiano et al.(2020)]{SanchezMenguiano2020}       
          S{\'a}nchez-Menguiano L., S{\'a}nchez S.F., P{\'e}rez I., et al., 2020, MNRAS, 492, 4149 

\bibitem [Schade et al.(1995)]{Schade1995} 
          Schade D., Lilly S.J., Crampton D., et al., 1995, ApJ, 451, L1 
          
\bibitem [Schaefer et al.(2020)]{Schaefer2020}  
          Schaefer A.L., Tremonti C., Belfiore F., et al., 2020, ApJL, 890, L3
          
\bibitem [Schaefer et al.(2022)]{Schaefer2022}  
          Schaefer A.L., Tremonti C., Kauffmann G., et al., 2022,ApJ, 930, 160 

\bibitem [Searle(1971)]{Searle1971}
          Searle L., 1971, ApJ, 168, 327

\bibitem [Sharda et al.(2021)]{Sharda2021}     
          Sharda P., Krumholz M.R., Wisnioski E., et al., 2021, MNRAS, 504, 53 

\bibitem [Sillero et al.(2017)]{Sillero2017}     
          Sillero E., Tissera P.B., Lambas D.G., Michel-Dansac L., 2017, MNRAS, 472, 4404 

\bibitem [Smith(1975)]{Smith1975}
          Smith H.E., 1975, ApJ, 199, 591
      
\bibitem [Speagle et al. (2014)]{Speagle2014}
          Speagle J.S., Steinhardt C.L., Capak P.L., Silverman J.D., 2014, ApJS, 214, 15

\bibitem [Tempel et al.(2018)]{Tempel2018} 
          Tempel E., Kruuse M., Kipper R., et al., 2018, A\&A, 618, A81
  
\bibitem [Torres-Flores et al(2020)]{TorresFlores2020} 
          Torres-Flores S., Amram P., Olave-Rojas D., et al., 2020, MNRAS, 494, 2785 
         
\bibitem [Vila-Costas \& Edmunds(1992)]{VilaCostas1992} 
          Vila-Costas M.B., Edmunds M.G. 1992, MNRAS, 259, 121

\bibitem [Vincenzo et al(2016)]{Vincenzo2016} 
          Vincenzo F., Belfiore F., Maiolino R., Matteucci F., Ventura P., 2016, MNRAS, 458, 3466 
          
\bibitem [Wilkinson et al.(2022)]{Wilkinson2022} 
          Wilkinson S., Ellison S.L., Bottrell C., et al.,  2022, MNRAS, 516, 4354
          
\bibitem [Williams et al.(2022)]{Williams2022} 
          Williams T.G., Kreckel K., Belfiore F., et al., 2022, MNRAS, 509, 1303

\bibitem [Wu et al.(2023)]{Wu2023} 
          Wu P.-F., Bezanson R., D'Eugenio F., et al., 2023, ApJ, 955,75

\bibitem [Yesuf et al.(2017)]{Yesuf2017} 
          Yesuf H.M., French K.D., Faber S.M., Koo D.C., 2017, MNRAS, 469, 3015
         
\bibitem [Zaritsky et al.(1994)]{Zaritsky1994}
          Zaritsky D., Kennicutt R.C., Huchra J.P., 1994, ApJ, 420, 87 
         
\bibitem [Zinchenko et al.(2016)]{Zinchenko2016} 
          Zinchenko I.A., Pilyugin L.S., Grebel E.K., S{\'a}nchez S.F., V{\'\i}lchez J.M., 2016, MNRAS, 462, 2016
                 
\end{thebibliography}
\end{document}